\documentclass[10pt]{article}

 \usepackage{amsmath,amsthm,amssymb}
 \usepackage{makeidx,epsfig,lscape}
 \usepackage{color,colortbl}
 \usepackage{fancyhdr}
 \usepackage{xcolor,pict2e}

 \renewcommand{\headrulewidth}{0pt}
 \renewcommand{\footrulewidth}{0.5pt}

 \definecolor{myaqua}{rgb}{0.0,0.5,0.55}
 \definecolor{lightaqua}{rgb}{0.75,0.95,0.95}

 \usepackage[colorlinks = true,
            linkcolor = myaqua,
            urlcolor  = blue,
            citecolor = myaqua]{hyperref}

\usepackage{floatrow}
\def\lin#1#2{\textcolor[rgb]{0.6,0.6,0.6}{\vspace*{#1mm} \hrule
   height 3 pt \vspace*{#2mm}}}
\def\bt{\begin{tabular}}
\def\et{\end{tabular}}
\def\and{\mbox{ and }}

\def\1{{\bf 1}}

 \def\sectionn#1{\refstepcounter{section}{\color{myaqua}

 \vskip 6mm

 \noindent\Large\bf\thesection. #1}

 \vskip 3mm}

 \def\boxx#1#2#3#4#5{
 {\linethickness{#4pt}\put(#1,#5){\color{myaqua}{\line(1,0){#3}}}}
 \multiput(#1,#2)(0,#4){2}{\line(1,0){#3}}
 \multiput(#1,#2)(#3,0){2}{\line(0,1){#4}}
  }

\begin{document}

 \fancyhead[L]{\hspace*{-13mm}
 \bt{l}{\bf Open Journal of *****, 2015, *,**}\\
 Published Online **** 2015 in SciRes.
 \href{http://www.scirp.org/journal/*****}{\color{blue}{\underline{\smash{http://www.scirp.org/journal/****}}}} \\
 \href{http://dx.doi.org/10.4236/****.2014.*****}{\color{blue}{\underline{\smash{http://dx.doi.org/10.4236/****.2015.*****}}}} \\
 \et}
 \fancyhead[R]{\includegraphics{pic1.ps}}

 $\mbox{ }$

 \vskip 12mm

{ % \fontfamily{Cambria}\selectfont

% "Title of the Paper"
{\noindent{\huge\bf\color{myaqua}
  Entropy, biological evolution and the  \\[2mm] psychological 
arrow of time}}
\\[6mm]
{\large\bf Torsten Heinrich{$^1$}, Benjamin Knopp{$^2$} and Heinrich P\"as{$^3$}}
\\[2mm]
{ %\fontfamily{Calibri}\selectfont
 $^1$ Institute for Institutional and Innovation Economics, University of Bremen, Germany     \\           
 Email: \href{mailto:torsten.heinrich@uni-bremen.de}{\color{blue}{\underline{\smash{torsten.heinrich@uni-bremen.de}}}}\\[1mm]
  $^2$ Fachbereich Physik, AG Neurophysik, Philipps-Universit\"at Marburg, Germany\\
  Email: \href{mailto:benjamin.knopp@physik.uni-marburg.de}{\color{blue}{\underline{\smash{benjamin.knopp@physik.uni-marburg.de}}}}\\[1mm]
   $^3$ Fakult\"at f\"ur Physik, Technische Universit\"at Dortmund,  Germany\\
   Email: \href{mailto:heinrich.paes@tu-dortmund.de}{\color{blue}{\underline{\smash{heinrich.paes@tu-dortmund.de}}}}\\[1mm]
 \\[4mm]
Received **** 2015
 \\[4mm]
Copyright \copyright \ 2014 by author(s) and Scientific Research Publishing Inc. \\
This work is licensed under the Creative Commons Attribution International License (CC BY). \\
\href{http://creativecommons.org/licenses/by/4.0/}{\color{blue}{\underline{\smash{http://creativecommons.org/licenses/by/4.0/}}}}\\
 \includegraphics{pic2.ps}

\lin{5}{7}

 { % \fontfamily{Cambria}\selectfont
 {\noindent{\large\bf\color{myaqua} Abstract}{\bf \\[3mm]
 \textup{
 We argue that in Universes where future and past differ only by the
entropy content a psychological arrow of time pointing in the
direction of entropy increase can arise from natural selection in
biological evolution.
We show that this effect can be demonstrated in very simple toy computer 
simulations of evolution in an entropy increasing or decreasing environment.
 }}}
 \\[4mm]
 {\noindent{\large\bf\color{myaqua} Keywords}{\bf \\[3mm]
 Arrow of Time; Time Asymmetry; Entropy; Evolution
}

 \fancyfoot[L]{{\noindent{\color{myaqua}{\bf How to cite this
 paper:}} T. Heinrich, H. P\"as and B. Knopp (2015)
 Entropy, biological evolution and the psychological 
arrow of time.
 ***********,*,***-***}}

\lin{3}{1}

\sectionn{Introduction}

{ \fontfamily{times}\selectfont
 \noindent 
 While the laws of classical physics are perfectly time-symmetric. i.e. 
causality works equally well forward and backward in time both daily life
and the microcosm governed by quantum laws are time-asymmetric
(excellent monographs on the nature of time include \cite{Zeh:1992vf,Price,Albert} 
and the collections of essays 
\cite{Halliwell:1994wq,Mersini-Houghton:2012mha}).
This asymmetry manifests itself in various arrows of time:

\begin{itemize}

\item
the thermodynamical arrow of time, as provided by the second law of 
thermodynamics: the entropy  in a closed system always increases 
or remains constant and
never decreases. For example, intact cups can fall off the table and splatter 
in thousand
pieces, releasing their potential energy as heat radiation, while
cup pieces on the floor never cool down their environment and use this 
energy to unbreak and jump onto the table.
Of course for a system to undergo entropy increase it has to
be out of equilibrium in the first place, i.e. starting from a low
entropy state.

\item
the cosmological arrow of time
\cite{Hawking:1985af,Kiefer:1994gp}: 
the Universe Hubble expands as time proceeds.

\item
the radiation arrow of time: (electromagnetic) waves propagate retarded, i.e. outward from the source.

\item
the quantum mechanical arrow of time: pure states `collapse' or decohere
(depending on one's preferred interpretation of quantum mechanics) into mixed states but mixed
states never evolve into pure states.

\item
the
CP asymmetry: weak interaction processes in particle physics proceed with different rates if the direction of time is reversed.

\item
the biological arrow of time: living creatures die but corpses do not revive.

\item
the psychological arrow of time: it is possible remember the past but not to
"pre-member" the future.

\end{itemize}
   
\renewcommand{\headrulewidth}{0.5pt}
\renewcommand{\footrulewidth}{0pt}

 \pagestyle{fancy}
 \fancyfoot{}
 \fancyhead{} % clear all header and footer fields
 \fancyhf{}
 \fancyhead[RO]{\leavevmode \put(-140,0){\color{myaqua}T. Heinrich, H. P\"as, B. Knopp} \boxx{15}{-10}{10}{50}{15} }
 \fancyhead[LE]{\leavevmode \put(0,0){\color{myaqua}T. Heinrich, H. P\"as, B. Knopp}  \boxx{-45}{-10}{10}{50}{15} }
 \fancyfoot[C]{\leavevmode
 \put(0,0){\color{lightaqua}\circle*{34}}
 \put(0,0){\color{myaqua}\circle{34}}
 \put(-2.5,-3){\color{myaqua}\thepage}}

 \renewcommand{\headrule}{\hbox to\headwidth{\color{myaqua}\leaders\hrule height \headrulewidth\hfill}}

It is interesting to note, though, that not all of these arrows of time are 
independent. Most obvious, the biological arrow of time is just another 
example of
entropy increase. 
After all, entropy in its statistical 
interpretation is nothing but a measure of how many microstates correspond
to a given macrostate, and thereby a measure of how probable this macrostate
is: there are simply many more possible ways for a cup to be broken
than to be unbroken, and many more ways for being dead 
%(like head chopped off, heart ripped out, etc...) 
than for being alive (almost all organs intact, working and in their right places).

Beyond that, in modern physics it actually becomes more and more common to assume that
{\it all} arrows of time are just different manifestations of entropy 
increase:

\begin{itemize}

\item 
the cosmological expansion may be intimately related to the thermodynamical 
arrow of time by actually {\it defining} the direction in which entropy
would increase by providing a low entropy state at the beginning of time. 
In such scenarios time would start to run backwards 
if the Universe would recollapse. Alternatively, the cosmological arrow may
be no arrow at all, as a Universe shrinking with time could be possible 
as well, given different initial conditions.

\item
the radiation arrow of time can result from boundary conditions as well.

\item
if the quantum mechanical arrow of time is due to decoherence \cite{Joos:1984uk}
as in the
Everett-Many-Worlds interpretation \cite{everett} it is simply a consequence of entropy 
increase once a quantum system is entangled with its environment in a 
measurement process. This results from the fact that the microstates of
the environment are not accessible to the observer.

\item
while the
CP symmetry is violated in weak interaction processes CPT remains conserved and may be interpreted as 
the correct operation of time inversion \cite{Carroll}.

\end{itemize}

The notion that time is an illusion, that it is nothing but entropy increase
after all, is further supported by the fact that the Wheeler-de-Wit equation~\cite{DeWitt:1967yk} 
\begin{equation}
\hat{H} | \psi \rangle = 0
\end{equation}
for the quantum mechanical wave function 
of the Universe \cite{Hartle:1983ai} in canonical quantum gravity models does not
include any time coordinate. In quantum cosmology, time may thus rather 
come into being as an emergent phenomenon when the transition to a semi-classical description of 
Nature is made, i.e. quantum fields evolving in a background of classical
gravity, since by separating out gravity as a classical background of spacetime
a macrostate is defined corresponding to
 many possible microstates in the full quantum 
gravity description (see e.g. the discussions in \cite{Zeh:1986ix,Barbour,Barbour:2009zd,Ellis:2008ar,Kiefer:2009tq,Rovelli:2009ee}).

\begin{figure}
\begin{minipage}[hbt]{6cm}
	\centering
	\includegraphics[width=6cm]{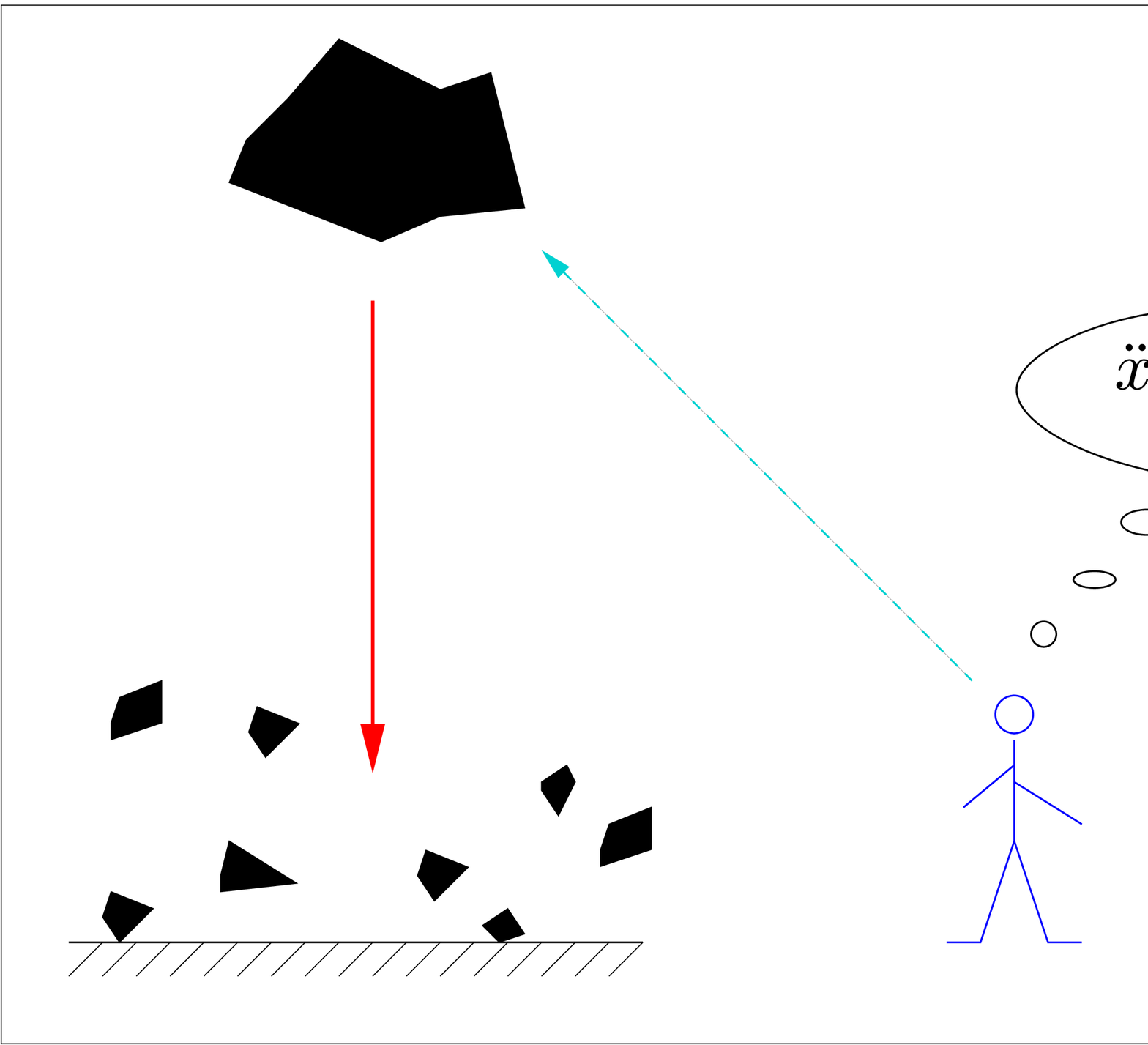}
\end{minipage}
\hfill
\begin{minipage}[hbt]{6.5cm}
	\centering
	\includegraphics[width=6.5cm]{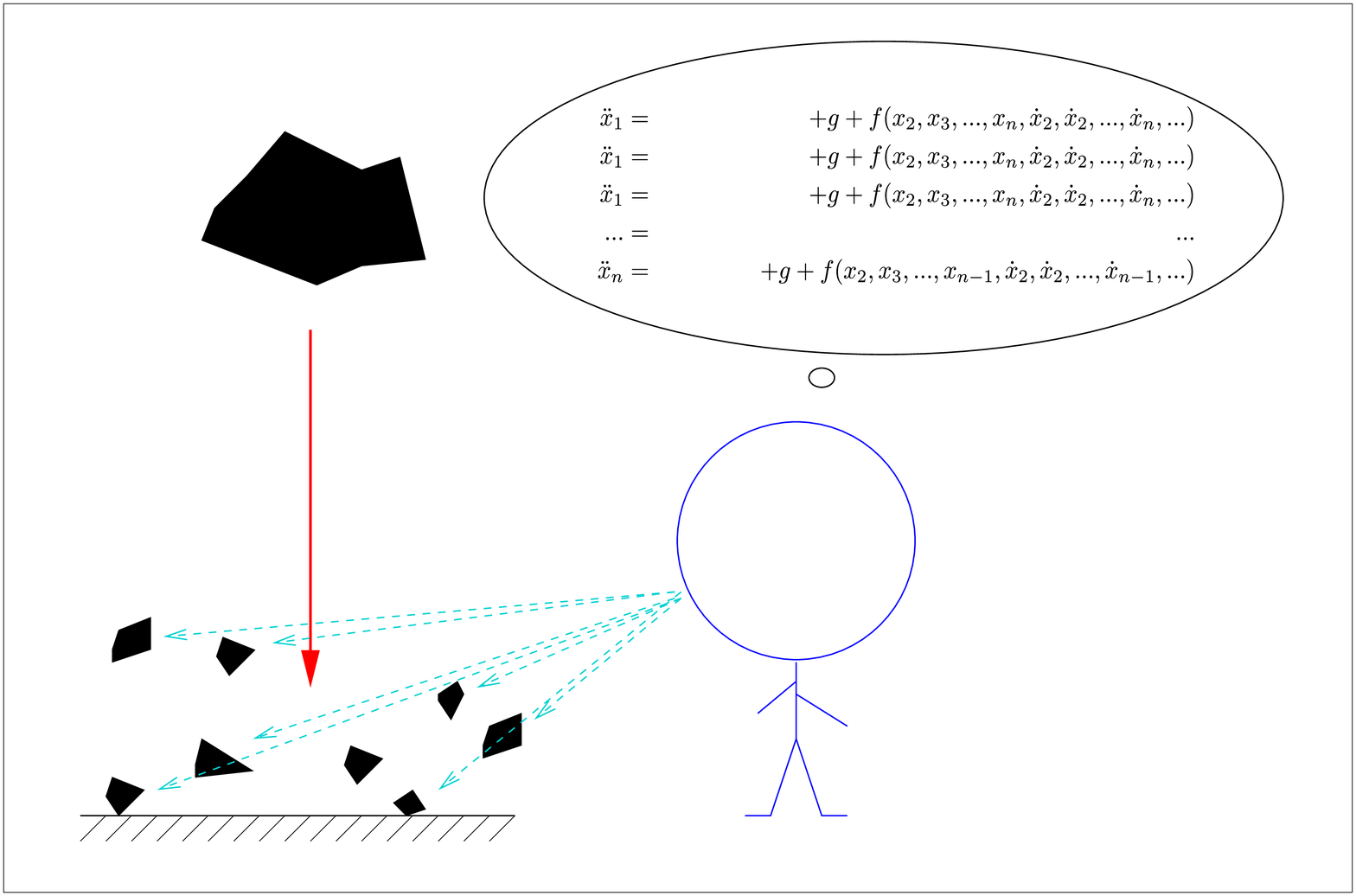}
\end{minipage}
\label{paul1}
\caption{
Illustration of an individual escaping 
a threat by processing 
memory stored about the past (left panel) and escaping
the same threat by processing memory stored about the future (right panel).
}
\end{figure}

In the following we thus adopt the hypothesis that entropy increase is the 
only manifestation of time. It should be stressed that this assumption makes time a 
psychological phenomenon. After all the entropy of a given macrostate depends on its definition
in terms of microstates, a property that can not be derived from the fundamental laws of physics.
If this hypothesis is correct, however, the most obvious manifestation of time becomes the most 
puzzling one:
the psychological arrow of time - the fact that we remember 
the past but do not "pre-member" the future.

More concrete, if the entropy budget is the only difference
between past and future one would expect that building up memory would 
correspond to entropy increase. Quite the opposite is true, though.
By organizing our brains in a way to store memory the unordered high entropy
state of the brain evolves into an organized low entropy state
\cite{landauer0,landauer1, landauer2, landauer3}
\footnote{
Note however that there exists a controversial exchange of arguments  between 
Hawking \cite{Hawking}, Hartle \cite{Hartle:2004wx}, Schulman \cite{Schulman} and Maroney \cite{Maroney}
on whether the psychological arrow can be explained as a special case of a
computational arrow of time
which has to proceed necessarily in the direction of entropy increase,
based on the argument that memory requires the erasure of record beforehand \cite{landauer0}.
We consider 
this controversy as unsettled.}. Of course this
is possible as the brain is an open system, and the process of building up
memory and reducing entropy locally inside the brain
can proceed at the cost of increasing entropy in the environment. 
But this consideration seems to suggest that if entropy increase is the only 
manifestation of time there would be no reason why we shouldn't pre-member
the future, thereby naturally growing entropy in our brain instead of forcing
a local entropy decrease by using up free energy.

In the following we will follow this line of reasoning and show that in the case the
fundamental laws of physics would allow to pre-member the future, 
natural selection in the course of biological evolution could eliminate
individuals developing memory of the future.

There exist at least two classical explanations for the psychological arrow of time
in the literature which are related to but different
from the approach adopted here (compare the discussions of Zeh \cite{Zeh:1992vf}, Albert \cite{Albert} and
Carroll \cite{Carroll} as well as 
\cite{Nikolic:1998iz,Mlodinow:2013eja}):
First, one can argue that memory is just a special case of a document
of the past. Such documents are more strongly correlated with events in the past than with
events in the future, since documents of the future could originate with a higher probability from a statistical fluctuation
(we will come back to this argument in the discussion section). 
Next, information is typically transmitted by (electromagnetic) radiation and thus subject to the radiation arrow of time 
which is a consequence of boundary conditions in an entropy increasing environment.    
One may argue that these explanations should be considered more fundamental than any explanation relying on the utility
of memory as they concern the basic concept and origin of memory itself, but this is not necessarily the case.
Recently, Rovelli argued for example that the coarse-graining hypothesis in defining a macrostate which is necessary to
calculate its entropy is a consequence of the interaction creatures "living in time" have with the Universe
\cite{Rovelli:2014cja,Rovelli:2015dha}.
Tegmark makes a similar argument when he supposes that the emergence of consciousness and the emergence of time may be related
\cite{Tegmark:2014kna}.
In this sense we adopt the Archimedean or atemporal standpoint demanded by Price \cite{Price} who argued that
that the asymmetry of causation is anthropocentric in origin. 
In the remainder of the paper we will thus focus on the superior predictive power of memory of the past as compared to memory 
of the future as a possible cause for the psychological arrow of time.

The basic argument here is simple. Memory has evolved in the course of 
biological evolution for only one reason: namely,
in order to increase the fitness of an individual
in a situation in the present. In its most basic form, memory is helpful
by providing a set of initial conditions which allow to calculate (or at 
least estimate) the behavior of enemies, prey or prospective sex partners
in the present. The fundamental laws of physics which govern this behavior 
are classical: they work as well forward as backward. 
However, in order to calculate a process in the present from initial 
conditions in the future one typically has to deal with a higher entropy 
state than by calculating the process from initial data in the past:
For an illustrative example, just imagine you try to avoid being hit by a rock. You see the
position of the rock on top of a hill and you can easily estimate how it will 
fall to the ground, where it disintegrates (see Fig.~\ref{paul1}, left panel). 
The problem reduces to solving a very simple differential equation.
If you want to achieve the same
goal by relying on initial data in the future, you have to know the positions
of all constituents of the disintegrated rock, and calculate how and where 
exactly they integrate to form the rock flying up onto the hill (Fig.~\ref{paul1}, right panel).
There are far more initial conditions to memorize, and the problem now transformed
into a complicated multi-dimensional set of coupled differential equations.

As entropy is a measure of the information required to describe the
microstate realized in a given macrostate, an individual would have to store and process
more information. And, depending on the situation it wants to manage, the
amount of information may not be a factor of few, but will typically grow
exponentially the farther your memory lasts into the future. Storing and
processing information requires storage, and - metaphorically oversimplified - 
the more storage one needs the
larger one's brain has to be (see e.g. the comparison of various animal brain sizes in \cite{fox}). 
At some point (and this point will come pretty 
soon) such a large brain will make the individual's ability to move and react 
(the only reason to develop memory in the first place) more and more difficult
and finally impossible. Such species will never prevail or even develop
in the course of evolution that favors agility and swift decision-making.%evolution triggered by the survival of the fittest.
%(of course an individual could gain energy by increasing the entropy budget of its brane
%but as the individual would still need much more energy to keep its lower vital 
%functions alive this benefit is not significant).
   
If we consider the psychological arrow of time as a possible consequence
of biological evolution though, we run into the difficulty that we typically 
understand evolution as a process {\it in time} while we now use it
to establish an (at least seemingly) very fundamental property of time
itself. The question is thus how we understand evolution and survival
in a time-symmetric Universe.

As we point out here, though,
it is not too difficult at all to imagine such a scenario. 
For sake of simplicity we model evolution by a computer simulation. 
A very simple realization would study two contrasting scenarios with 
one in which the lifeforms remember the past while in the other they 
pre-member the future\footnote{
Note that the model does not really rely on any "flow of time" at all (as demanded by Huw Price 
for an atemporal perspective \cite{Price}) but can be considered as different states linked by a causal
connection (the physical laws) without explicit reference to time at all.
}.

\sectionn{A simple toy model}
{ \fontfamily{times}\selectfont
 \noindent 
In order to address the question whether remembering populations have a superior
fitness compared to pre-membering populations we have developed a very simple 
simulation %cellular automaton \cite{neumann} 
as a toy model for artificial life and evolution of a population in a 
hostile environment.

The individuals of a given population are modeled as agents 
%by the states of cells 
moving in
a two-dimensional grid.
A hostile environment is simulated by adding falling rocks with a given mass 
which move predictably according to the laws of classical physics.
%just like billiard balls bouncing off the walls of the box.
At the boundary of the grid the rocks disintegrate into two rocks with half of the mass
of the original rock. The exact rules for collisions with the boundaries are depicted in
Fig.~\ref{rock-movements}. Once the mass of a rock reaches the value of 0.5 it constitutes 
no threat for individuals anymore and is removed from the simulation.
New rocks are generated randomly with a given probability in order to ensure a stable environment 
with a more or less constant rock density oscillating around an average value.
The rocks are responsible for both %natural 
selection and for the irreversibility of processes in the 
environment which allows to define the direction of time as the direction of increasing entropy.  
The evolution of the rock environment is being recorded first and can subsequently be run both
in forward and backward direction. Finally populations of individuals can be placed in either
the forward or the backward running environment and their probabilities for extinction or survival %survival fitness 
can be compared.

\begin{figure}[!t]
\centerline{
\includegraphics[width=0.8\textwidth]{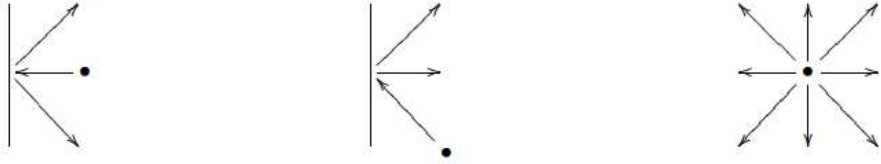}
}
\caption{
Rock movements and disintegration on the walls of the grid.
}
\label{rock-movements}
\end{figure}

The individuals possess a given number of lives which gets reduced by the mass value of 
a rock when they get hit.
When the number of lives is reduced to zero the individual "dies" and gets removed from the
simulation.
The individuals reproduce asexually by reduplication with a defined probability $p_r$ with each time step. %in a given time interval.
Furthermore the individuals scan the adjacent and next-to-adjacent cells (a two-cell Moore-neighborhood) 
of their respective position and avoid a threatening rock by moving out of its way once it is detected.
The limited memory of an individual is modeled by allowing individuals to find and avoid only one single rock -
whenever two or more rocks threaten the individual it is at danger of being hit.
If no rock is observed the individuals move or rest randomly. For a screen shot of the simulation see Fig.~\ref{screen-shot}.

\begin{figure}[!t]
\centerline{
\includegraphics[width=0.9\textwidth]{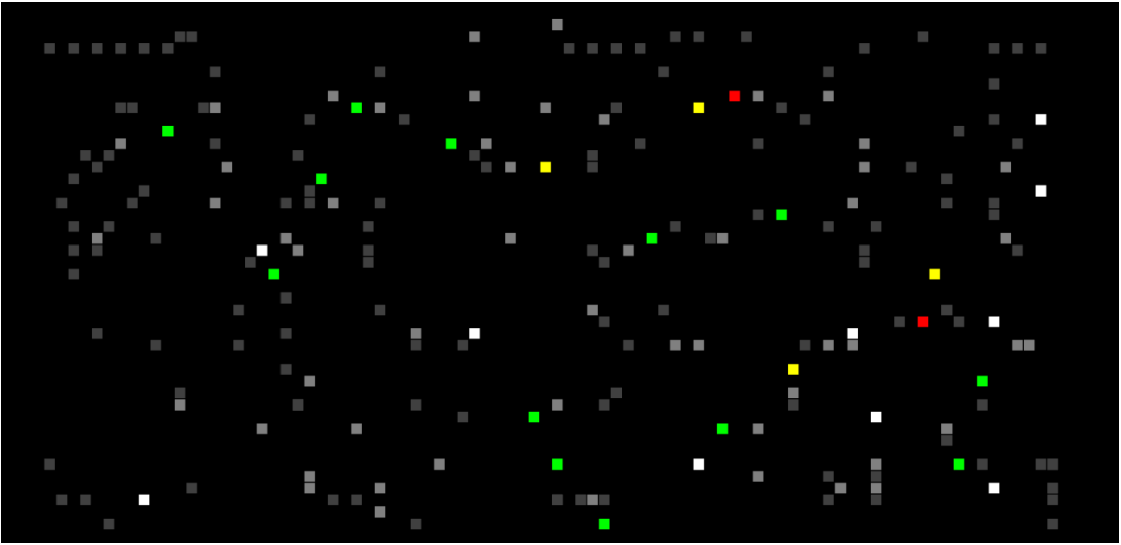}
}
\caption{
Screen shot of the simulation. Individuals are depicted as green (3 lives), yellow (2 lives) or red (1 live) cell,
rocks are depicted as white (mass 4), light grey (mass 2) or dark grey (mass 1) cells.
}
\label{screen-shot}
\end{figure}

%The individuals execute their free will by maximizing the distance to the next
%rock. 
%``Sex and violence'' are then simulated by the simple rules
%which apply on encounter of two individuals of the same type or of
%one or more individuals ($I$ )with a predator/rock ($R$) in the same pixel:
%\begin{eqnarray}
%I + I &\rightarrow& 3~I \\
%I + R &\rightarrow& 0. 
%\end{eqnarray}
%As it is easy to fathom
%how the simulation will evolve
%we can run the program as a Gedankenexperiment. As the dynamics of the 
%predators/rocks are deterministic and consequently are the maximization
%algorithms for the movement of individuals, we can run the program 
%both forward as well as backwards in time. For the first species the 
%lifespan is defined as the time span from the appearance of the first
%set of individuals to the annihilation of the last individual by a 
%predator/rock. Analogously the rules can be defined for the 2nd species,
%just reverese in time. One can run both simulations separately and will
%find a longer lifespan for species 1 living forward in time moving faster.

\sectionn{Results}
{ \fontfamily{times}\selectfont
 \noindent 
In order to compare the probabilities for extinction or survival %survival fitness 
of past-remembering and future-premembering populations a population
of individuals is situated either in the forward or backward running environment of scattering rocks and
evolved according to the simple rules discussed above. As a consequence, populations can remain stable over
a given time span, die out or explode. 
%The statistical fluctuations due to the random creation of rocks and random
%replication of individuals result in all possible outcomes even for a single set of initial conditions. 
For high $p_r$ or a low number of stones, populations are more likely
to explode, for low $p_r$ or high numbers of stones, they are more
likely to become extinct. In between there is a critical region. Due
to statistical influences induced by the random creation of rocks and
random replication of individuals as well as their random initial
positions the simulation may result in all possible outcomes even for
a single set of initial conditions within the critical region. To
study the exact differences between the two settings we investigate
(past remembering and future pre-membering lifeforms), 
we thus performed a 
statistical analysis, defining the value $p_{1/2}$ as the reproduction probability $p_r$ for which more than half of
the simulations result in a stable population and survive a given running time of 5000 iterations (for an example of a 
stable population over a certain time span see Fig.~\ref{stable-population}).\footnote{Note that $p_{1/2}$ may vary with the length of the simulation, particularly for smaller numbers of iterations. For the number of iterations used here (5000) this effect is greatly reduced. One might consider using the threshold for which the probability for the population to grow or to decline would be equal. However, this would result in many simulation runs with - potentially exponentially - expanding numbers of agents which, in turn, would increase the number of computation operations per iteration accordingly leading to requirements in computation power that is by magnitudes larger than what we can employ.}

\begin{figure}[!t]
\centerline{
\includegraphics[width=1.0\textwidth]{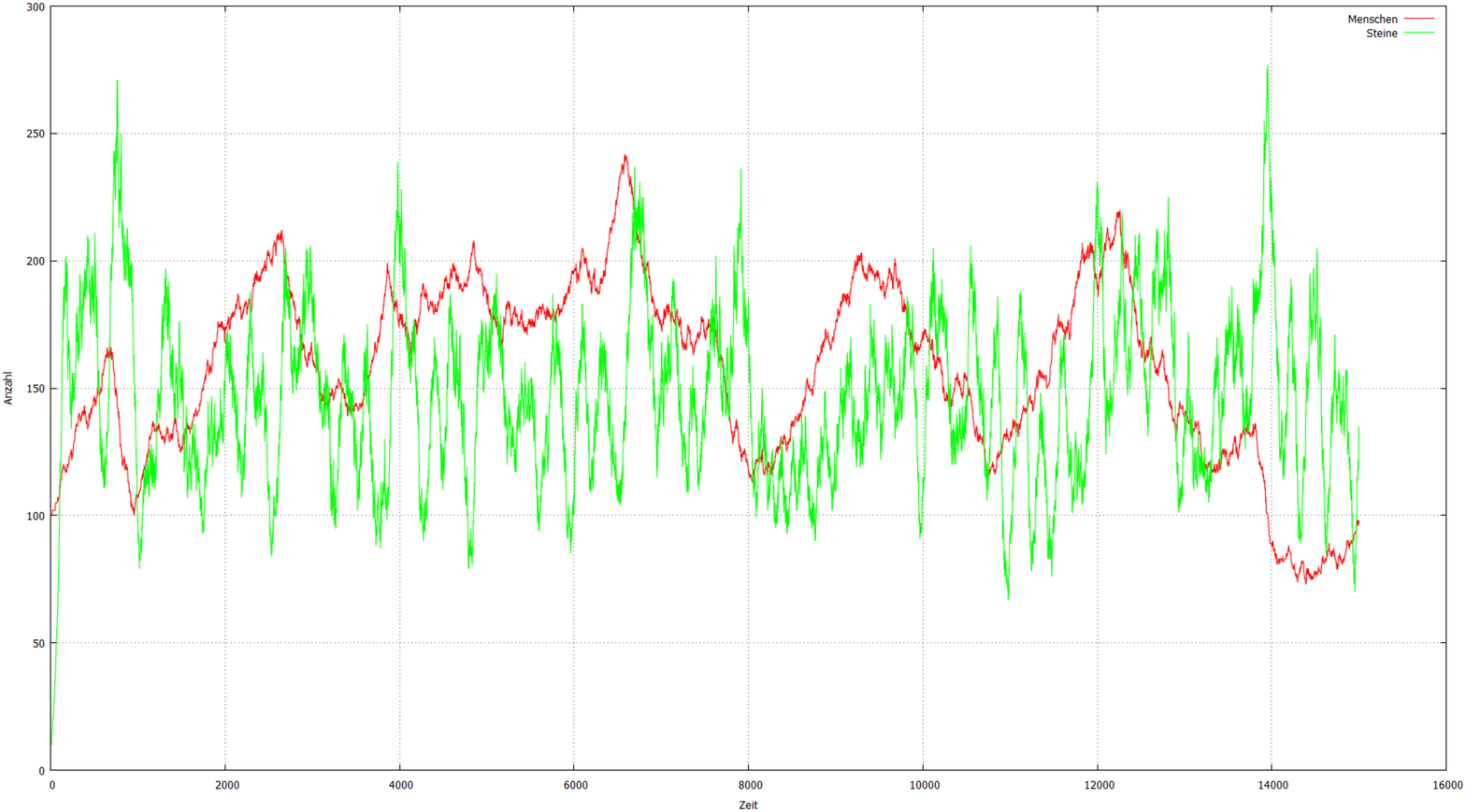}
}
\caption{Stable population of individuals (red) in comparison to the number of rocks in the grid (green) as a function of running time/iterations.
}
\label{stable-population}
\end{figure}

The result is shown in Fig.~\ref{result} where
the percentage of stable populations is plotted as a function of $p_r$.
Here red and green crosses correspond to past-remembering and future-premembering populations, respectively and 
the respective values for $p_{1/2}$ can be read off as a result of linear extrapolations to the 50\% survival rate.

\begin{figure}[!t]
\centerline{
\includegraphics[width=1.0\textwidth]{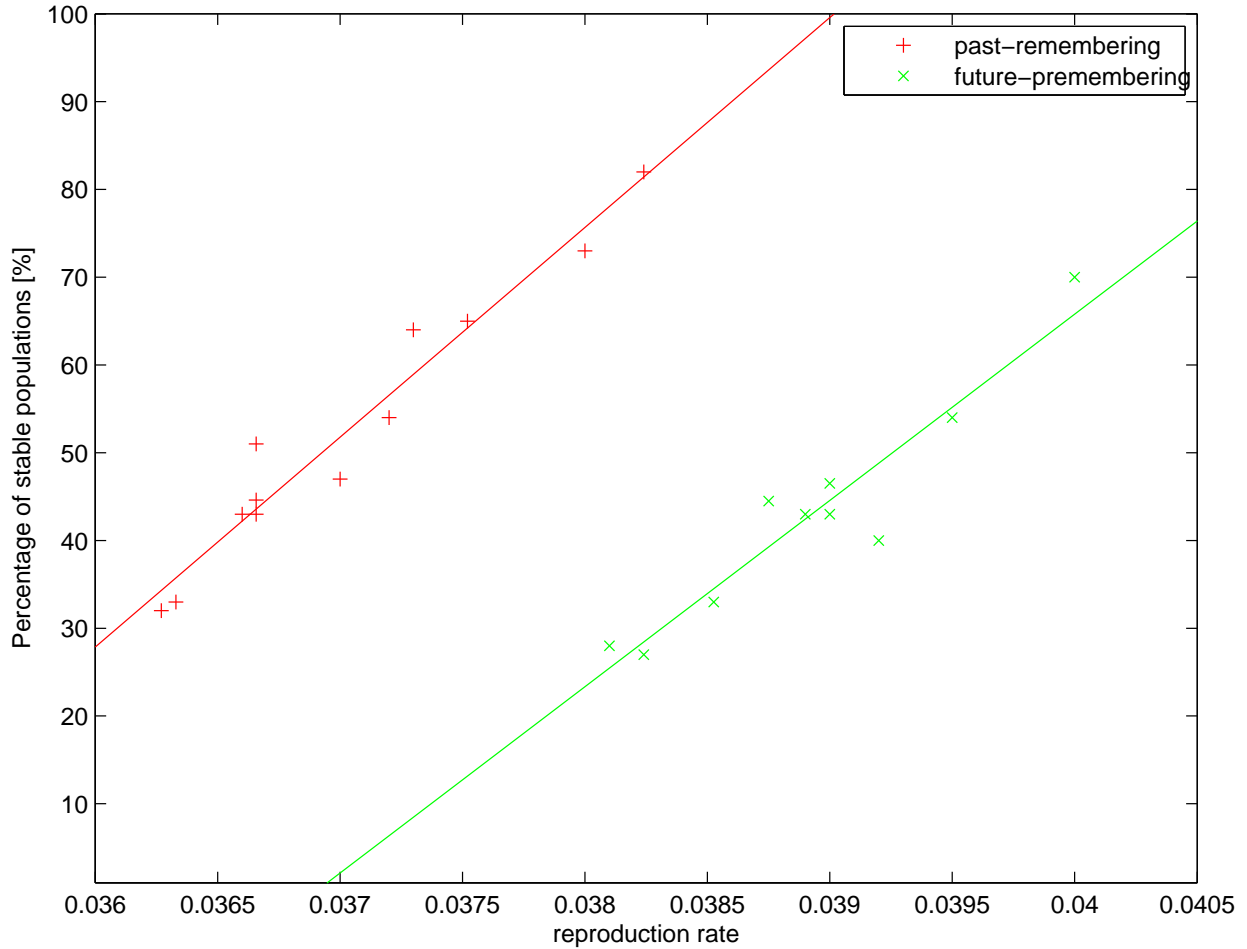}
}
\caption{
Percentage of stable populations as a function of the reproduction rate $p_r$ for past remembering (red crosses, upper fit)
and future-premembering populations (green crosses, lower fit). The mean reproduction rate with a 50 \% survival chance $p_{1/2}$
is obtained by linear extrapolation. 
}
\label{result}
\end{figure}

For populations in an entropy increasing environment the reproduction rate necessary for a stable population
is $(3.7 \pm 0.002) \cdot 10^{-2}  $ while for populations in an entropy-decreasing environment the necessary rate is 
$(3.9 \pm 0.009) \cdot 10^{-2}$.\footnote{This difference and its statistical significance seems to be stable across different parameter settings (while the values themselves are of course sensitive to parameter changes). The parameter setting we used includes an initial number of agents of $30$, an initial number of rocks of $10$ with an initial mass of $3$ each and a probability of $0.4$ that a new rock is injected in each iteration on a rectangular grid of $43 \times 23$ cells. The simulation runs had a length of $5000$ iterations; the simulations for each candidate $p_r$ were repeated 100 times in order to obtain a good estimate for the corresponding survival rate.}

%Of course a more realistic simulation would include both species in
%one single environment. This can be realized easily as well.
%While it looks confusing on first glance that in this
%picture individuals of the future-premembering species seem to be created by rocks
%and annihilated in inverse reduplication, the respective lifespan is still
%an adequate measure of evolutionary success. 
%A more serious challenge is that for the future-premembering species the initial time  
%where its
%population is set free
%would lie in the future.  
%In order to avoid this problem one can introduce creation from mutation 
%by allowing
%individuals to appear spontaneously at any time. 
%Success in natural selection 
%is then defined by the number of individuals of one species present averaged 
%over the running time of the simulation. Again, the faster species will
%excel in avoiding its threats and thus be more abundant on average.
%This effect will be even more pronounced if some competition for 
%resources like food between the past-remembering and future-premembering species is introduced, e.g. by 
%making the probability for reproduction 
%dependent on the number of individuals of the competing species.

\sectionn{Discussion}
{ \fontfamily{times}\selectfont
 \noindent 
At this point a few remarks are in order:

First, we chose to adopt in our simulation what may be called a 
{\it psychological perspective},  as it 
corresponds to the subjective perception of the pre-membering population 
(starting with birth and ending with death).
That means, if we define the direction of time as the direction of 
increasing entropy, 
our pre-membering populations are not really 
living forward in time with a memory of the future, but
rather live entirely backwards in time. 
We chose the psychological perspective here in order to have a simple,
model that allows to compare re-membering and pre-membering populations
in a transparent way. 
It could be argued though that the
former, {\it biological} perspective is more realistic as life
typically is understood as a process evolving in the direction of
entropy increase. 
As the mechanisms responsible for the time asymmetry are the 
same in both cases we do not expect a qualitative change in our results, 
if the biological perspective is chosen instead. 

Such a biological perspective has been employed in a simulation 
performed by Zeissner \cite{Zeissner}, where a previously recorded environment 
could be accessed by remembering and pre-membering agents at times $t_0-1$
and $t_0+1$, respectively. It was noted that under these conditions the pre-membering
population could access the same information as the remembering population, albeit two time
steps earlier. This results in an advantage of the pre-membering population which competes with
the disadvantage of the minor predictive power of the "rediction" process. Which effect dominates depends on the
actual situation but it seems clear that for realistic environments the superior predictive power
of past-memory will be more important.

Second, in our simulation
the cost of information storage in the memory is not modeled. 
We rather compared the capabilities of populations to survive in entropy 
increasing and entropy decreasing environments assuming they have the same 
skills in information processing.
This can be motivated by the fact that the brain is an open 
system, and that recording memory in our brains works against the 
external entropy increase anyway, but it leaves open the question whether 
backward living lifeforms can store information at no or even negative cost.
As a psychological arrow of time requires to sustain the physical basis
for memory storage (a ``big brain'') over a large time period in parallel to the 
biological arrow of time corresponding to entropy increase though we consider any possible 
amount of free energy generation by storage of pre-membrance as negligible.

Third, as discussed previously and related to the point above, it has been pointed out by 
Zeh \cite{Zeh:1992vf},
Albert \cite{Albert},
Carroll \cite{Carroll} and others, that if memory can be generated 
automatically at no cost in free energy, ``false memory'' in the sense of 
information stored inside the brain which has no correspondence to real information
about the environment 
is more probable and thus such memory will be less reliable. This effect will
strengthen the evolutionary argument for back-in-time remembrance but
is more difficult to simulate and probably less important than the simple
argument advocated here.

\sectionn{Conclusions}
{ \fontfamily{times}\selectfont
 \noindent 
In this paper we were studying a very simple toy model modeling biological
evolution of past-remembering and future-premembering populations
in order to compare the probabilities for extinction or survival %survival fitness 
of these populations.
While this work obviously does not prove that the real psychological arrow of time
is a product of biological evolution and relies on many assumptions,
is extremely simple and custom built,   
it does demonstrate, though, that --
if the fundamental laws of physics would allow to pre-member the future --
natural selection could be sufficient to eliminate this skill. 

A critical argument often raised is that the simulation compares only backward memory
with future memory, but not with hybrid memory concepts, typically phrased as "a limited
knowledge of the future could be very helpful". This argument misses the fact that any kind of memory
is costly. This is the reason why we don't remember the entire past but only an apparently most
significant subset of it in the first place. Taking the extreme case, already by remembering the entire past a deterministic Universe
would allow Laplace's demon to calculate the entire future with absolute precision at a lower or equal cost than by pre-membering
the future. This argument suggests that {\it any} useful amount of memory of the future would be too costly to prevail
in a natural selection process.

Finally, a well-motivated question is how to test these ideas. More realistic computer simulations
beyond the simple toy model described here would
definitely help. More convincing of course would be to probe the idea
in an actual biological experiment. 
%Species like Drosophila seem complex 
%enough to have some kind of memory of the past allowing them to react 
%properly on threats or chances in the present, while at the same time they 
%provide  a role model for genetics due to fast reproduction.
In principle one could study for example the evolution of a population of bacteria 
in a local environment where entropy is always decreasing.
If the argument advocated in this work is correct it should be possible for such a 
population to develop memory of the future, but it is totally unclear whether such
a development could occur fast enough to be observable and if
the population could survive long enough under such conditions to evolve sufficiently.\footnote{Note that ``memory'' here does not refer to the genome of these bacteria 
%since this would be the same as with past-remembering lifeforms and would not function in the same way backwards in time. Instead, memory of the future might occur in other, 
but rather to
less complex information processing mechanisms the bacteria rely on, for instance the presence or absence of chemical substances in the bacterium which plays a part in governing its movement in species that are able to actively move through bacterial gliding.}
To discuss how realistic such a setup is thus beyond the scope of this work.
 
In summary we
thus consider this work as a proof of principle that it is possible to discuss
a possible origin of the psychological arrow of time from natural selection
in a meaningful scientific way.

 {\color{myaqua}

 \vskip 6mm

 \noindent\Large\bf Acknowledgments}

 \vskip 3mm

{ \fontfamily{times}\selectfont
 \noindent
 HP would like to thank Peter Schupp for inspiring his research in this direction
by arguing the psychological arrow of time may actually result 
from psychology, as well as 
Marc Brown, Sean Carroll, Lou Clavelli, Frank Deppisch, 
Carlo Rovelli, Ketil Soerensen, R\"udiger Vaas, Sinan Zeissner, Heinz-Dieter Zeh for useful discussions during various stages of this work.

 {\color{myaqua}

}}

\end{document}